\documentclass[fleqn,10pt, twocolumn]{wlscirep}
\usepackage{setspace}
\usepackage{textcomp}
\usepackage{amsmath,amsthm, amssymb}
\usepackage{bm} 

\usepackage{braket}
\usepackage{siunitx}
\usepackage[position=top]{subfig}
\usepackage{stackengine,graphicx,subfig}

\title{Quantum cryptography with highly entangled photons from semiconductor quantum dots }

\author[1,*]{Christian Schimpf}
\author[1]{Marcus Reindl}
\author[1]{Daniel Huber}
\author[1]{Barbara Lehner}
\author[1]{Saimon F. Covre Da Silva}
\author[1]{Santanu Manna}
\author[2]{Michal Vyvlecka}
\author[2]{Philip Walther}
\author[1,*]{Armando Rastelli}

\affil[1]{Institute of Semiconductor and Solid State Physics, Johannes Kepler University, Linz 4040, Austria}
\affil[2]{Vienna Center for Quantum Science and Technology, Faculty of Physics, University of Vienna, Austria}

\affil[*]{Corresponding authors: christian.schimpf@jku.at}

\begin{document}
\maketitle

\textbf{
State-of-the-art quantum key distribution systems are based on the BB84 protocol  \cite{Bennett1984} and single photons generated by lasers. These implementations suffer from range limitations \cite{Lucamarini2018} and security loopholes, which require expensive adaptation \cite{Lo2005}. The use of polarization entangled photon pairs \cite{Ekert1991,Bennett1992,Acin2007} substantially alleviates the security threads while allowing for basically arbitrary transmission distances when embedded in quantum repeater schemes \cite{Briegel1998,Kimble2008,BassoBasset2019,Zopf2019}. Semiconductor quantum dots are capable of emitting highly entangled photon pairs \cite{Huber2018} with ultra-low multi-pair emission probability \cite{Schweickert2018} even at maximum brightness. Here we report on the first implementation of the BBM92 protocol \cite{Bennett1992} using a quantum dot source with an entanglement fidelity as high as \SI{0.97(1)}{}. For a proof of principle, the key generation is performed between two buildings, connected by a \SI{350}{m} long fiber, resulting in an average key rate of \SI{135}{bits/s} and a qubit error rate of \SI{1.91}{\percent} over a time span of 13 hours, without resorting to time- or frequency-filtering techniques. Our work demonstrates the viability of quantum dots as light sources for entanglement-based quantum key distribution and quantum networks. By embedding them in state-of-the-art photonic structures, key generation rates in the Gbit/s range are at reach.}

Successful quantum key distribution experiments with entangled photons (EQKD) have been carried out \cite{Jennewein2000,Yin2020}, utilizing photon pairs generated via the spontaneous parametric down conversion (SPDC) process. However, for those sources, the multi-photon-pair emission probability is directly coupled to the source brightness by their Poissonian emission characteristics. This circumstance limits the pair-extraction efficiency for SPDC sources to about \SI{0.01}{}, as higher values inevitably increase the average photon-pair number \cite{Jons2017}. The excess photons lead to spurious detector clicks during the key generation in EQKD protocols, which results in key errors and security loopholes \cite{Beaudry2008}. Semiconductor quantum dots (QDs) can generate polarization-entangled photon pairs \cite{Benson2000, Akopian2006, Dousse2010, Juska2013} and do not suffer from these limitations due to their sub-poissonian photon-pair emission characteristics \cite{Benson2000, Senellart2017}. GaAs quantum dots obtained by droplet etching \cite{Gurioli2019} can emit polarization entangled photon pairs with a demonstrated multi-photon emission probability as low as \SI{8(2)E-5}{} \cite{Schweickert2018}. The weak confinement of the multi-particle wave functions in these QDs \cite{Huber2019} in combination with their high in-plane symmetry result in entanglement fidelity values as high as \SI{0.97(1)}{} (see below), without resorting to time-filtering (as used in Ref. \cite{Dzurnak2015}) or post-growth tuning \cite{Huber2018}. The high fidelity values allow for a low qubit error rate (QBER) and are therefore crucial for the viability and the effective key rate of EKQD implementations \cite{Bennett1992,Bennett1992Crypt,Acin2007,Gisin2002}. Successful attempts of quantum teleportation \cite{Reindl2018} and entanglement swapping with QDs \cite{BassoBasset2019,Zopf2019} further encourage the development of QD-based quantum networks for long-haul quantum communication \cite{Kimble2008}.\\
\begin{figure*}[ht]
    \centering
    \includegraphics[width=17.5cm]{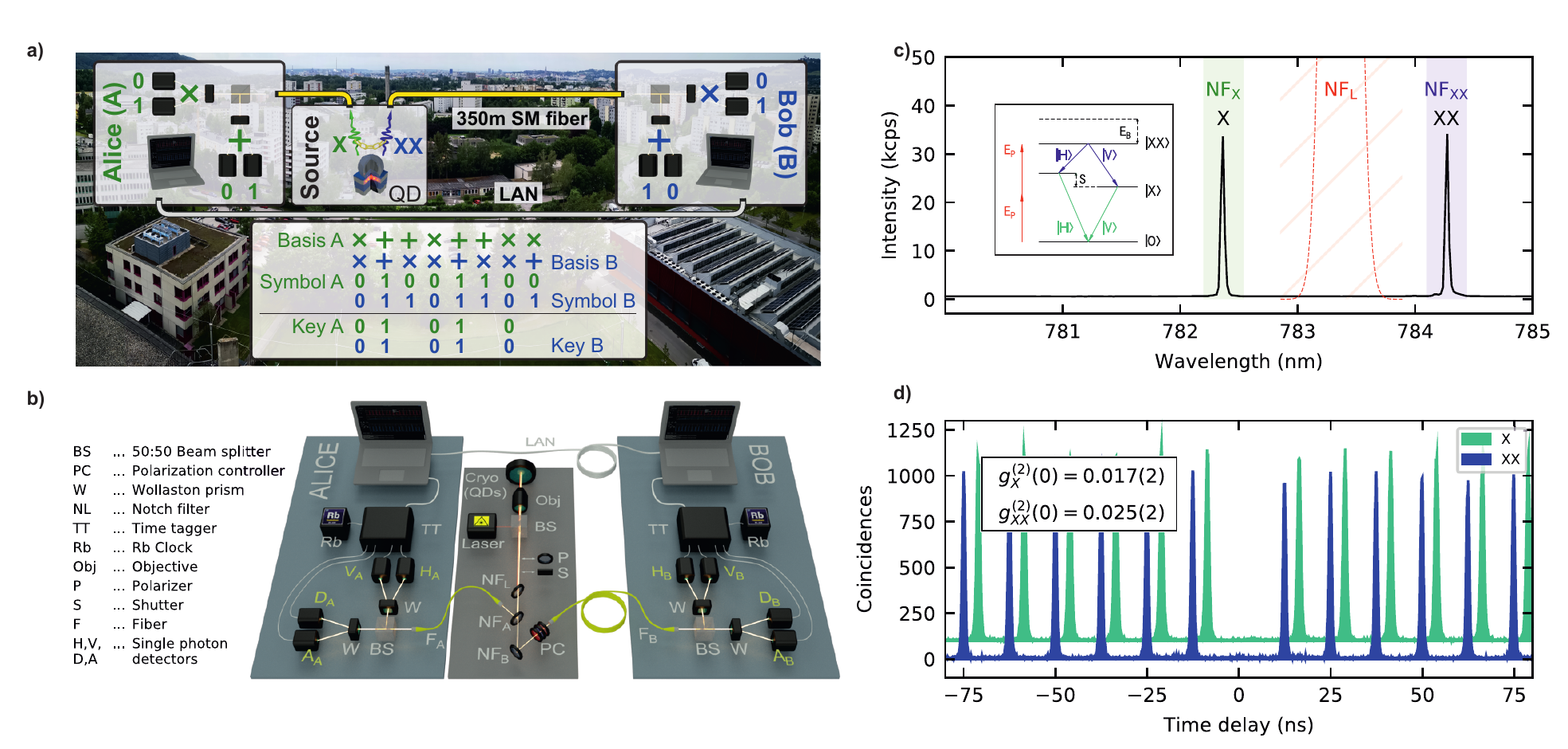}
\caption{(a) Configuration of the fiber-based EQKD system using highly entangled photons from a GaAs quantum dot (QD) photon source. Alice is situated in the lab, together with the photon source. Bob is assembled in a movable box placed in the LIT Open Innovation Center and connected to the photon source via a \SI{350}{m} long single mode (SM) fiber. The photons are analyzed in the rectilinear ("+") and diagonal ("$\times$") basis, each outcome is assigned to a "0" or a "1" symbol. The key sifting, as sketched in the table, is performed over a \SI{100}{MBit/s} LAN connection. (b) Illustration of the optical and electronic components of the setup. The GaAs QD light source is in a He flow cryostat and is resonantly excited by a pulsed laser with a repetition rate of \SI{80}{MHz}. A set of identical NFs reject laser stray light and distribute the entangled photons through $F_\text{A}$ and $F_\text{B}$ to Alice and Bob. The PC corrects for any reversible polarization altering effects. Two identical four-state analysis setups at Alice and Bob measure the photons in rectilinear and diagonal basis. The TTs are connected to standard computers and backed by Rb clocks, acting as a \SI{10}{MHz} frequency reference. A movable linear polarizer (P) and a shutter (S) are used for initial clock synchronization. (c) Emission spectrum of the resonantly driven QD after rejecting the laser stray light by NF\textsubscript{L} and selecting the X and XX lines by NF\textsubscript{X} and NF\textsubscript{XX}, respectively. The inset sketches the two-photon-excitation scheme (TPE), with $E_P$ the laser photon energy, $S$ the fine structure splitting, and $E_B$ the binding energy of the $\ket{\text{XX}}$ state. (d) Autocorrelation of the X and XX signals, using a time bin of $\SI{1}{ns}$.}
\label{fig:OIC}
\end{figure*}

The layout of our EQKD implementation is depicted in Fig. \ref{fig:OIC}(a). The first communication node ("Alice") and the GaAs QD-based photon source are situated in a laboratory in the semiconductor physics building at the Johannes Kepler University campus, while the second node ("Bob") is a mobile system placed on an office desk in the LIT Open Innovation Center (OIC) and is connected to the source by a \SI{350}{m} long single mode (SM) fiber. Entangled photon pairs are distributed via fibers from the source to Alice and Bob. A more detailed depiction of the setup is shown in Fig. \ref{fig:OIC}(b). The BBM92 protocol relies on analyzing the incoming photons in the rectilinear ("+") and diagonal ("$\times$") basis \cite{Bennett1992}. A perfectly random choice of the measurement basis plays an essential role for the security of the protocol. We exploit the natural unpredictability of the path taken by a photon impinging on a 50:50 beam splitter (BS) to ensure this randomness. Different techniques can be employed to reduce the number of detectors, but at the cost of higher effort to ensure the randomness of the basis and a higher vulnerability to side channel attacks \cite{Dzurnak2015,Jennewein2000}. Each detector after each of the two outputs of the BS (the two possible bases) is assigned either a symbol "0" or "1". Alice and Bob perform the key sifting procedure by exchanging only information about the measurement basis via a standard \SI{100}{MBit/s} LAN connection, leaving no information for a potential eavesdropper on the public channel. Only if the entangled photon pairs arriving at the detectors of Alice and Bob are in the maximum entangled polarization state $\ket{\phi^+}$ with respect to the measurement basis, photons analyzed in the same basis cause the same measurement outcome at both sides and therefore identical key strings consisting of "0" and "1" symbols.

In order to generate entangled photon pairs, the QD is initialized into the biexciton state by a pulsed laser, using a two-photon excitation scheme (TPE) \cite{Huber2018}, see inset of Fig. \ref{fig:OIC}(c). Figure \ref{fig:OIC}(c) shows the spectra of the consecutively emitted photons from the biexction-to-exciton (XX) and the exciton-to-groundstate (X) radiative decay for a selected QD. As we drive the XX state in a resonant process at saturation (at the so-called "$\pi$ pulse" condition), a pair emission occurs with a probability of $\epsilon=\SI{0.87(5)}{}$ per pump pulse. Figure \ref{fig:OIC}(d) shows the autocorrelation histograms for the XX and X signals under these pumping conditions after filtering the remaining laser stray light. The resulting multi-photon emission probabilities of $g^{(2)}_{\text{X}}(0)=\SI{0.017(2)}{}$ and $g^{(2)}_{\text{XX}}(0)=\SI{0.025(2)}{}$ are predominantly governed by the APD dark counts and ambient light.  These values emphasize the excellent single-photon purity of GaAs QDs in combination with the TPE scheme and can be further improved by a better shielding against stray light and by using detectors with lower dark counts~\cite{Schweickert2018}. For SPDC sources, in contrast, the multi-photon emission probability is directly connected to the pair-emission probability by the Poissionian emission characteristics, so that $g^{(2)}_{\text{SPDC}}(0)=\epsilon_{\text{SPDC}}$ \cite{Orieux2017}. Hence, efficient pumping and a high single photon purity inherently exclude each other in the case of SPDC sources.

The maximum excitation pump rate for QD sources is only limited by the XX and X radiative lifetimes $T_{\text{1,XX}}$ and $T_{\text{1,X}}$, as no excitation can occur until the QD has relaxed into the groundstate. For the here used photonic structures ($T_{\text{1,XX}}\approx\SI{120}{ps}$ and $T_{\text{1,X}}\approx\SI{270}{ps}$) a pump rate of about \SI{600}{MHz} is accessible (see supplementary for calculations). Reducing the lifetimes further by Purcell enhancement \cite{Liu2019} can readily push this limit to the GHz regime.\\
One of the most important parameters affecting the fidelity to the $\ket{\phi^+}$ state for QD sources is the so-called fine structure splitting (FSS) between the two X eigenstates, which, for the used QD, has a magnitude of $S=\SI{390(62)}{neV}$ (A detailed description the optical setup, TPE and the relation of the FSS with the entanglement is given in the methods).

A practical realisation of EQKD requires distribution of entanglement over noisy quantum channels, like the here used single mode fibers. The goal is to distribute the entangled photon pairs while ensuring the highest possible fidelity to the $\ket{\phi^+}$ Bell state. The first aspect to be considered in this case is the synchronization between Alice and Bob. The latter requires precise knowledge about the photon's arrival times at the nodes, which are possibly separated by large distances and exposed to different environmental conditions. Their hardware and internal clocks therefore inevitably operate dissimilarly. In the methods, we describe in detail how we handle the timing and synchronization between Alice and Bob with external Rb clocks, and also propose an alternative without reference clocks. An important core feature of the here used synchronization method is the irrelevance of the employed hardware, the fiber length and the latency times, as it relies solely on the strong polarization correlation between the emitted XX and X photons.

The second aspect to be considered is the influence of the fibers themselves on the entangled photons. The most pronounced effect of single mode (SM) fibers on propagating light is a general damping of the signal, which is about \SI{3}{dB/km} at a wavelength of $\SI{780}{nm}$. Besides damping, more profound effects on entangled photons have to be taken in account when distributed over fibers: polarization dependent loss (PDL) \cite{Huttner2000} and polarization mode dispersion (PMD)  \cite{Antonelli2011}. PDL does not depend on the source's coherence properties and is less of an issue for modern fibers at length scales at which damping becomes the dominating limiting factor anyway (at about $\SI{200}{km}$ for \SI{1550}{nm} or $\SI{10}{km}$ for \SI{780}{nm}). PMD, on the other hand, not only randomly (but reversibly) alters the polarization state, but can lead to irreversible degradation of the entanglement \cite{Antonelli2011,Lim2016}. Sources with a photon coherence time $T_2$ lower than the fibers' differential group delay (DGD) (typically around $\SI{0.5}{ps/\sqrt{km}}$) suffer substantially from PMD and one has to resort to frequency filtering \cite{Poppe2006}. Although strategies exist to preserve the entanglement through PMD- and/or PDL-afflicted quantum channels \cite{Jones2018}, they require precise knowledge about the effects and control about the polarization modes simultaneously, rendering them impractical in the context of a real-world application due to the constantly changing environment. In the case of QDs, the $T_2$ time is capped by $T_{\text{1,XX}}$ and $T_{\text{1,X}}$, which are orders of magnitudes higher than the DGD of SM fibers and therefore PMD has no meaningful impact on the entanglement (see supplementary for supporting calculations).

The complex rotation of the polarization state induced by the fibers and the rest of the setup is reversible and can be cancelled by a polarization controller (PC). As long as the variation of the rotation (e.g. due to temperature fluctuations) is slow compared to the measurement time, it can be eliminated effectively during the whole key generation procedure by readjusting the PC on demand, see methods. For testing the capabilities of the PC and to probe the entanglement fidelity, preliminary experiments were performed, where Bob was placed together with Alice in the lab and connected by a \SI{700}{m} long fiber (\SI{350}{m} to the OIC and \SI{350}{m} back to the lab). After the PC, the fidelity to the $\ket{\phi^+}$ state, measured by a full state tomography, was determined as $f_{\ket{\phi^+}}=\SI{0.97(1)}{}$. Details to the measurement, including the 2-qubit density matrix, are given in the supplementary.


\begin{figure*}[ht]
    \centering
    \includegraphics[width=17.5cm]{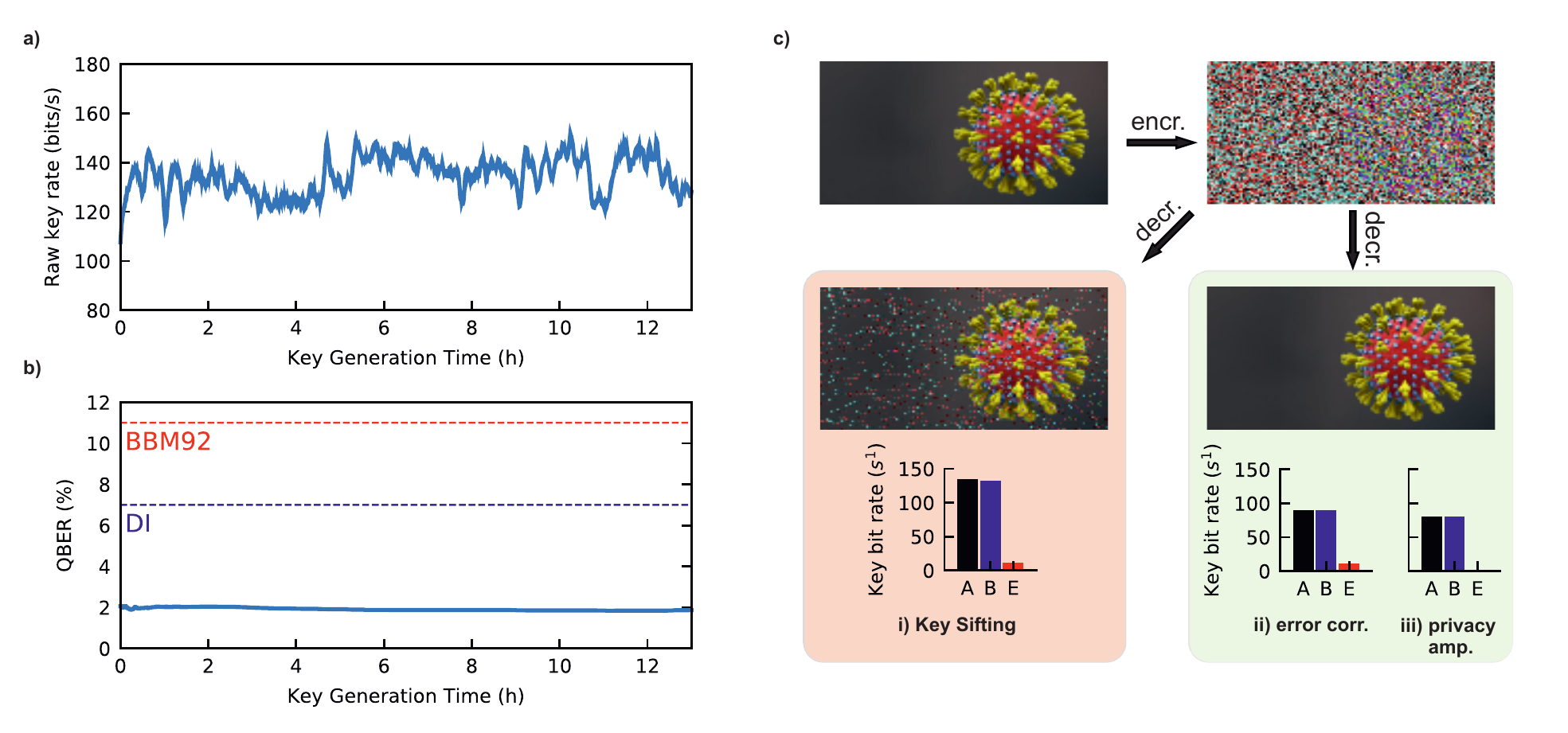}
\caption{(a) Raw key rate and (b) QBER over a time span of 13 hours. The dashed lines indicate the upper limits of \SI{11}{\percent} for the BBM92 and \SI{7}{\percent} for the Device Independent (DI) protocols. (c) Exemplary encryption (encr.) of a \SI{29.2}{kByte} large bitmap with Alice's key using the one-time-pad method, with a subsequent decryption (decr.) with Bob's key. The decrypted bitmaps are shown for the cases (i) directly after key sifting and (ii,iii) after error correction and privacy amplification. The bar charts indicate the distribution of information over Alice (A), Bob (B) and the eavesdropper (E) after each step.}
\label{fig:Crypto}
\end{figure*}

After the initial polarization correction, the setup is prepared for continuous key generation. During this operation, the software executed at Alice's and Bob's Notebooks permanently monitors the QBER by comparing a random and subsequently discarded subset of the generated keys. The random choice prevents a potential attacker from finding predictable time windows for eavesdropping without being detected. During normal operation, only a small fraction of keys has to be sacrificed for this purpose to achieve appropriate accuracy and security. However, the target here was to record the QBER with a high frequency and with high accuracy, so we dedicated \SI{10}{\percent} of the keys and excluded this fraction from the key rate calculations. For the key generation, photons within an arrival time bin of \SI{1}{ns} are used, which includes all photons from the QD, but excludes a reasonable amount of background photons from ambient light. Figure \ref{fig:Crypto}(a) shows the raw key rate with an average of $\SI{135}{bit/s}$ over a time span of 13 hours. The monitored QBER during the key generation is shown in Fig. \ref{fig:Crypto}(b). The dashed lines indicate the maximum QBER of \SI{11}{\percent} for the here employed BBM92 \cite{Bennett1992,Gisin2002} and \SI{7}{\percent} for the Device Independent (DI) \cite{Acin2007} protocols, respectively, below which it is still possible to distill secure keys. The DI protocol protects against potential collective attacks and includes an additional measurement basis at one node, which can be readily implemented in the present system. Without adjusting the PC during the measurement, we observed a variation of the QBER between \SI{1.84}{\percent} and \SI{2.06}{\percent} over the course of the measurement, probably stemming from slight changes of the fiber's environment, such as temperature and vibrations. The resulting average raw QBER is \SI{2.36}{\percent}. The discrepancy between QBER of \SI{1.1}{\percent} calculated from the measured entanglement and the here measured minimum of \SI{1.84}{\percent} is the result of several minor imperfections, which add up to a significant degree (see methods for details). Compared to preliminary tests with short fibers (\SI{2}{m}) in the lab we observed no increase in average QBER, as expected due to the negligible effect of PDL and PMD on the photons from QDs.

The fraction of "1" symbols in the resulting key strings is $\SI{0.513}{}$, which reveals a slight bias towards the "1" symbol. This discrepancy arises mostly from slight differences of the efficiencies of the single photon detectors in the four-state measurement arrangements. The resulting non-unity Shannon entropy poses a minor potential security threat, but can be corrected by inverting a random half of the bits for both keys \cite{Acin2007}. The finite QBER can be brought to zero during the act of error correction by iterative parity exchange \cite{Bennett1992Crypt,Jennewein2000} and subsequent "bit twiddling" \cite{Bennett1988}, where both procedures leak a negligible portion of information about the keys to the public channel. Assuming an unknown measurement apparatus, one always has to expect the initial finite QBER to stem from an intercept-resend attack from an eavesdropper tampering with the quantum channel \cite{Gisin2002}. The number of leaked deterministic bits - plus five standard deviations - can be estimated \cite{Bennett1992Crypt} by $l= 4Nq + 5\sqrt{12Nq}$, where $N$ is the length of the key and $q$ the QBER. In order to erase this information, a "privacy amplification" technique \cite{Bennett1988} is applied: The key is compressed to a maximum length of $N-l$ by using a random universal hash function on which Alice and Bob agree over the public channel. Figure \ref{fig:Crypto}(c) depicts the encryption of a bitmap using \SI{29.2}{kByte} of Alice's key (acquired in about \SI{46}{min}), utilizing the information-theoretically secure one-time-pad method. The latter is performed by applying the bit-wise XOR operation of the original bitmap and Alice's key. The decryption is done by performing the same operation with the encrypted message and Bob's key. The bar charts in (i-iii) depict the distribution of information about the key among Alice (A), Bob(A) and a potential eavesdropper (E) after key sifting, error correction and privacy amplification, respectively. The latter decrease the net key rate depending on the QBER (see supplementary for calculations), hence a small QBER is desired for a high secure key rate. The average key rate achieved here with the \SI{350}{m} long fiber is \SI{135}{bits/s} before and \SI{86}{bits/s} after correction with the initial QBER of \SI{1.91}{\percent} and a final QBER of 0.

In conclusion, we have presented an implementation of the BBM92 \cite{Bennett1992} quantum key distribution protocol using semiconductor quantum emitters as sources of highly entangled photon pairs with an ultra-low multi-photon emission probability. With polarization entangled photons from droplet etched GaAs quantum dots (QDs) \cite{Huber2018} embedded in a simple planar cavity, a raw key rate of \SI{135}{bits/s} and a QBER of $\SI{1.91}{\percent}$ have been achieved through a \SI{350}{m} long optical fiber without time filtering, resulting in a net secure key rate of \SI{86}{bits/s}. These values allow for implementing protocols with a higher demand on the QBER, like the device independent scheme \cite{Acin2007}.

The long coherence time of photons emitted by QDs prevents a degradation of the polarization entanglement when transported via single mode fibers over practically relevant distances of hundreds of kilometers \cite{Antonelli2011}. The general damping in fibers, however, currently limits the transmission rates for the wavelength used here (about \SI{780}{nm}). Although free space entanglement distribution at this wavelength was successfully performed over a distance of \SI{1120}{km} \cite{Yin2020}, utilizing the well established fiber-based telecom infrastructure is desirable for the targeted, highly interconnected quantum networks. To this end, entangled photons in the telecom band (around \SI{1550}{nm}) could be obtained either by down-converting \cite{Weber2019} photons emitted by the GaAs QDs used here or by further developing QDs based on different materials \cite{Olbrich2017,Xiang2019}.

To increase the source brightness, the QDs can be embedded into circular Bragg resonators (CBR), which have demonstrated pair extraction efficiencies over \SI{0.6}{} \cite{Liu2019} and Purcell enhancements of up to 11 \cite{Wang2019}.  Using these structures, key rates of Gbit/s - well beyond the capabilities of parametric-down-conversion sources - are at reach. Considering the simultaneously high photon indistinguishably \cite{Liu2019}, QDs have the potential to act as nodes in much desired quantum networks \cite{Reindl2018,BassoBasset2019,Kimble2008}.

\small

\bibliography{main}


\section*{Methods}
\subsection*{Sample growth and photonic structure}
The GaAs QDs employed here were grown by the local droplet etching method by molecular beam epitaxy. The QDs are embedded in a \SI{124}{nm} thick $\text{Al}_{0.33}\text{Ga}_{0.67}\text{As}$  planar lambda-cavity, sandwiched between two distributed Bragg reflectors (DBRs). The DBRs are composed of 9 (bottom) and 2 (top) pairs of $\text{Al}_{0.20}\text{Ga}_{0.80}\text{As}$ and $\text{Al}_{0.95}\text{Ga}_{0.05}\text{As}$ layers with \SI{60}{nm} and $\SI{69}{nm}$ thickness, respectively. With an additional solid immersion lens (SIL) on top, this yields an extraction efficiency of $\eta_e\approx\SI{0.11}{}$.

\subsection*{Resonant two-photon excitation}
The sample is placed in a He flow cryostat, cooled to $T=\SI{5}{K}$. Resonant two-photon excitation (TPE) is performed using wavelength tunable laser pulses with a repetition rate of \SI{80}{MHz}, a pulse-width of about \SI{10}{ps} and a pulse-energy of about \SI{15}{fJ} (average pulse power of $\SI{1.1}{\mu W})$, which are focused on a single QD. By a resonant two-photon absorption process, almost each laser pulse excites the QD to the biexciton state, leading to a photon pair emission probability per laser pulse of $\SI{0.87(5)}{}$. This value can be extracted from X/XX cross-correlation measurements, as shown in the supplementary.

\subsection*{Photon collection and separation}
The emitted XX and X photons collected by the objective (Obj in Fig. \ref{fig:OIC}(b)) pass a set of notch filters NF\textsubscript{L} with a bandwidth of \SI{0.2}{nm} each, which reflect the largest part of the excitation laser stray light. The NF's central wavelength can be tuned from $\SI{780}{nm}$ to $\SI{786}{nm}$ by adjusting their tilting angle. Two NFs of the same type are individually tuned to reflect only the X (NF\textsubscript{A}) and XX (NF\textsubscript{B}) photons, which are then coupled into the single mode fibers F\textsubscript{A} (leading to Alice) and F\textsubscript{B} (leading to Bob). Depending on the experiment, the length of F\textsubscript{B} can vary between \SI{2}{m} and \SI{700}{m}.

\subsection*{Four-state measurement}
Alice and Bob both analyse their incoming photons by two nominally identical four-state measurement apparatus, which form the core of the BBM92 arrangement. In each setup a 50:50 beamsplitter (BS) is used to randomly select the measurement basis by directing the photons in one of two arms of the setup. In the reflected path, a Wollaston prism (W) is rotated such that one of its eigenaxis is approximately parallel to the optical table plane, which we define as the $\ket{H}$ polarization. As a consequence, the photons are measured in the rectilinear basis $\{\ket{H},\ket{V}\}$ by the following avalanche photodiods (APDs) $H_{A,B}$ and $V_{A,B}$, connected to a time tagger (TT), which registers the detector clicks. As a side-effect of this particular arrangement the rectilinear basis forms an eigenbasis of the polarization transformation caused by the BS reflection, which can therefore safely be neglected in the following considerations (see prove in supplementary). In the transmission path of the BS, the W is rotated by $\SI{45}{\degree}$ with respect to $\ket{H}$, so that the signal is measured in the diagonal basis $\{\ket{D},\ket{A}\}$ by the APDs $D_{A,B}$ and $A_{A,B}$.

\subsection*{Spectroscopy and autocorrelation measurement}
The spectra of the X and XX signals are analyzed by a standard reflective diffraction spectrometer with a resolution of about $\SI{30}{\mu eV}$. The fine structure splitting, despite being below the spectrometers resolution, can be determined by polarization mapping (see supplementary for details). The autocorrelation of the X or XX signal can be obtained by combining the detector events from $H,V$ and $D,A$, respectively, in order to mimic a Hanbury Brown and Twiss setup. No subtraction of the background is performed, which is predominantly defined by the APD dark counts. The values of $g^{(2)}(0)$ are obtained by calculating the peak area at zero time delay within a time bin of \SI{1}{ns} and comparing it to the average side peak areas.

\subsection*{Polarization entangled two-photon state}
The photon-pair's polarization state is described by\\ 
$\ket{\psi(t)}_{\text{X,XX}}=\frac{1}{\sqrt{2}}\left(
    \ket{H^\prime}_\text{X}\ket{H^\prime}_\text{XX} + e^{-\frac{i}{\hslash}S\,t}\ket{V^\prime}_\text{X}\ket{V^\prime}_\text{XX}
    \right)$,\\
where $S$ is the magnitude of the FSS between the two possible bright exciton eigenstates, $t$ is the time the QD dwells in the exciton state before decaying, and $\ket{H^\prime}$ and $\ket{V^\prime}$ are two orthogonal linear polarization states, defined by the in-plane-symmetry of the QD. The entangled state undergoes a time dependent phase evolution, determined by the product of $S$ and $t$. As no time filtering is employed the two-qubit density matrix is composed by all possible time dependent states weighted by their emission probability:
$
\rho_S = \int\limits_{0}^{\infty} \dfrac{1}{T_{1,\text{X}}} e^{-\frac{t}{T_{1,\text{X}}}}
\ket{\psi(t)}\bra{\psi(t)}\, \text{d}t$,
where $T_{1,\text{X}}$ is the exciton lifetime. Non-perfect single photon-pair emission and background from stray light and detector dark counts result in a non-unity single photon purity $\kappa=1-1/2 \left(g^{(2)}_{\text{X}}(0)+g^{(2)}_{\text{XX}}(0)\right)$, generalizing the equation to $\rho= \kappa\,\rho_S + (1-\kappa)\frac{I^{(4)}}{4}$, with $I^{(4)}$ the $4\times 4$ identity matrix. The here used and most common quantity for evaluating the similarity of a mixed state $\rho$ to a pure state $\ket{\phi^+}$ is the fidelity $f_{\ket{\phi^+}}(\rho) := \text{Tr}(\rho \, \ket{\phi^+}\bra{\phi^+})$.\\
The weak confinement of the biexciton and exciton wavefunctions within the GaAs QD result in short lifetimes of about $T_{1,\text{XX}}=\SI{120}{ps}$ and $T_{1,\text{X}}=\SI{250}{ps}$. The distribution of $S$ over different QDs shows values typically below $\SI{5}{\mu eV}$, i.e. QDs with $S<\SI{1}{\mu eV}$ are straightforward to find. This leads to values of $f_{\ket{\phi^+}}$ of over $\SI{0.95}{}$ without resorting to strain tuning.

\subsection*{Timing and synchronization}
The internal clocks of the time taggers (TT) are backed up by Rb clocks, acting as a \SI{10}{MHz} frequency standard. The detection time delay between the X and XX photons can be found by exploiting the strong polarization correlation between the entangled X and XX photons. Once determined, the time delay can be tracked continuously during QKD operation without leaking information about the key on the public channel. See supplementary for a detailed explanation.

\subsection*{Polarization control}
For cancelling the polarization altering effects induced by the fibers in an optimal and efficient manner for the given purpose, an heuristic approach was chosen: We assume the static transformations $\hat{F_A}$ and $\hat{F_B}$ in the two-dimensional polarization space for the fibers F\textsubscript{A} and F\textsubscript{B}, respectively. The entangled state undergoes the bi-local transformation to $\ket{\psi^{\prime}(t)}= \left( \hat{F_A} \otimes \hat{F_B} \right)\ket{\psi(t)}$. If the transformation is purely governed by PMD, $\hat{F_A}$ and $\hat{F_B}$ are unitary and the following equation holds (see supplementary for proof): $\ket{\psi^{\prime}(t)}= \left( \hat{1} \otimes \hat{F_B}\hat{F_A}^{\dagger} \right)\ket{\psi(t)}$, where $\hat{1}$ is the unity operator. This implies that the combined transformations of both fibers can be cancelled by inducing an additional unitary transformation $\hat{U}$ anywhere along X or XX light path, so that $\hat{U}\hat{F_B}\hat{F_A}^{\dagger}=\hat{F_B}\hat{F_A}^{\dagger}\hat{U}=\hat{1}$. A polarization controller (PC in Fig. \ref{fig:OIC}(b)), consisting of three rotatable wave plates (two quarter wave plates and one half wave plate) is capable of generating an arbitrary unitary transformation $\hat{U}(\bm{\theta})\in \text{SU(2)}$ in the polarization space (see supplementary for details), where $\bm{\theta}=(\theta_1,\theta_2,\theta_3)$ represents the three rotation angles. An optimum for $\bm{\theta}$ is found by observing the correlation between Alice's and Bob's detectors and minimizing the coincidences in the orthogonal bases $O_i \in \{H_A V_B, V_A H_B, D_A A_B, A_A D_B\}$ (later corresponding to key errors) while maximizing the coincidences in the colinear bases $C_i \in \{H_A H_B, V_A V_B, D_A D_B, A_A A_B\}$ (later corresponding to valid key entries), which is equivalent to minimizing the loss model function\\
$
    L(\bm{\theta}) := \sum \limits_{i=1}^{4} \braket{O_i|\rho(\bm{\theta})|O_i}+
    \left( \frac{1}{2}- \braket{C_i|\rho(\bm{\theta})|C_i} \right)
$,
with $\rho(\bm{\theta})$ the two-photon density matrix for the rotation angles $\bm{\theta}$. A downhill-simplex optimization algorithm is utilized for minimizing $L$ in about 6 minutes in average when starting without initial guess. The qubit error rate (QBER) is then given by $q = \frac{1}{4}\sum \limits_{i=1}^{4} \braket{O_i|\rho|O_i}$.\\
In the supplementary we provide data for an exemplary polarization correction using an effective fiber length of \SI{700}{m}.

\section*{Acknowledgments}
Christian Schimpf is a recipient of a DOC Fellowship of the Austrian Academy of Sciences at the Institute of Semiconductor Physics at Johannes Kepler University, Linz, Austria.\\
This work was financially supported by the Austrian Science Fund (FWF): P 29603, P 30459, I 4320, the Linz Institute of Technology (LIT) and the LIT Lab for secure and correct systems, supported by the State of Upper Austria\\
We thank C. Diskus for providing the Rb clocks, S. Zeppetzauer for his assistance in the lab, and J. Handsteiner and R. Kueng for the fruitful discussions.




\end{document}